\documentclass[prl,twocolumn,showpacs,floatfix,superscriptaddress,amsmath,amssymb]{revtex4-1}
\usepackage{graphicx}
\usepackage{dcolumn}
\usepackage{bm}
\usepackage{natbib}
\usepackage{color}

\newcommand{\ba}{\begin{eqnarray}}
\newcommand{\ea}{\end{eqnarray}}
\newcommand{\nn}{\nonumber}
\newcommand{\dk}{\delta_{1}}
\newcommand{\dkk}{\delta_{2}}
\newcommand{\dkkk}{\delta_{3}}
\newcommand{\dnk}{\delta_{n}}

\def\be{\begin{equation}}
\def\ee{\end{equation}}
\def\n{\nonumber\\}

\begin{document}

\title{Anharmonic effects in magnetoelastic chains}

\author{D.C.\ Cabra}
\affiliation{
Departamento de F\'{\i}sica, Universidad Nacional de la Plata,
C.C.\ 67, (1900) La Plata, Argentina
~~\\
Facultad de Ingenier\'\i a,  Universidad Nacional de Lomas de
Zamora, Cno. de Cintura y Juan XXIII, (1832) Lomas de Zamora,
Argentina.}

\author{C.J.\ Gazza}
\affiliation{
Facultad de Ciencias Exactas, Ingenier\'\i a y Agrimensura,
~~\\
Universidad Nacional de Rosario and Instituto de F\'\i sica Rosario,
~~\\
Bv. 27 de Febrero 210 bis, 2000 Rosario, Argentina}

\author{C.A.\ Lamas}
\affiliation{
Departamento de F\'{\i}sica, Universidad Nacional de la Plata,
C.C.\ 67, (1900) La Plata, Argentina}

\author{H.D.\ Rosales}
\affiliation{
Departamento de F\'{\i}sica, Universidad Nacional de la Plata,
C.C.\ 67, (1900) La Plata, Argentina}

\date{\today}
\begin{abstract}
We describe a new mechanism leading to the formation of rational magnetization plateau phases,
which is mainly due to the anharmonic spin-phonon coupling. This anharmonicity produces plateaux
in the magnetization curve at unexpected values of the magnetization without explicit magnetic frustration
in the Hamiltonian and without an explicit breaking of the translational symmetry.
These plateau phases are accompanied by magneto-elastic deformations which are not present in the harmonic case.
\end{abstract}

\maketitle

\section{Introduction}

Coupling of electronic and elastic modes has been shown to play a
crucial role in many condensed matter systems, most notably in the
BCS theory of superconductivity where the presence of the lattice
degrees of freedom is crucial to explain pair formation \cite{BCS}. Another
paradigmatic case is the so-called Peierls effect, where modulations
in the charge or spin densities may appear due to the electron-phonon
interactions (See {\it e.g.} \cite{reviewSP} and references therein).
More recently, phonon effects have been observed in many
other strongly correlated systems, in particular in some magnetic systems
which show plateaux in their magnetization curves \cite{MilaSrCuB}.

Usually one expects to have an accurate description of an
electron-phonon system by approximating the phonon potential
with a quadratic function of the interatomic distances between nearest
neighbor ions on sites $i$ and $j$, $\delta_{ij}$. Within the
same degree of accuracy, the dependence in $\delta_{ij}$
of the hopping amplitudes and/or the magnetic exchange constants is
approximated as a linear function.
This description works well in most of the cases since interatomic displacements
are usually rather small as has been verified experimentally in many
systems, like in the BCS superconductors.
More recently however, a less conventional BCS superconductor, MgB$_2$,
has shown an unusually high critical temperature, around 40$^0$K, which
could be the consequence of strong anharmonicities both in the phonon
potential and in the electron-phonon coupling \cite{NNMZA,LMK,Yildirim,AR}.

The relevance of anharmonic couplings has also been discussed in relation
to a great variety of compounds, both from an experimental
\cite{experiments1,experiments2,experiments3} and a theoretical point of view \cite{anharmtheory},
including the family of pyrochlore oxide superconductors, AOs$_2$O$_6$ for A=Cs, Rb,
and K \cite{experiments1}, the heavy fermion superconductors PrOs$_4$Sb$_{12}$,
SmOs$_4$Sb$_{12}$ \cite{experiments2}, and some potentially thermoelectric materials
such as X$_8$Ga$_{16}$Ge$_{30}$ (X=Eu, Sr, Ba) \cite{experiments3}, etc.
Another possible relevance of anharmonicities is in the study of spin systems
in high pulsed magnetic fields and Raman experiments \cite{HR}.

Apart from possible experimental motivations, the role of anharmonicities in
the physics of low dimensional systems is interesting in its own right and we
investigate this issue in the present paper in one of the simplest and most
paradigmatic one-dimensional systems: the $XXZ$ Heisenberg chain.

More precisely, we analyze in the present paper the
effects of anharmonic (adiabatic) phonons in the spin-Peierls mechanism as well
as the consequences on the magnetic properties of
the $XXZ$ Heisenberg chain coupled non-linearly to
lattice deformations. The most important consequence of the anharmonicity is that it
produces plateaux in the magnetization curve at unexpected values of the magnetization.
For example a plateau at $M=1/3$ of saturation magnetization appears without explicit
magnetic frustration in the Hamiltonian
and without an explicit breaking of the translational symmetry \cite{plateaux}, \cite{untercio}.
Besides, magnetoelastic deformations appear in some particular cases with frequencies
which halve that of the first harmonic, $2k_F$, as {\it e.g.} at $M=1/5$ (see below).
Similar conclusions should apply to more complicated models, since the
effects of other interactions such as {\it e.g.} a next-nearest neighbor interaction
would be simply to enlarge the extension of the plateaux phases and to modify
the magnitude of the spin gaps \cite{us}.

\section{The Model}

We start from the following spin-phonon
Hamiltonian in the limit of large ionic mass $M \rightarrow \infty$, the
so-called adiabatic limit
\ba
{\cal H}&=& J\sum_{i} (1+A_1\delta_{i}+A_2
\delta_{i}^2)\, \vec{S}_{i} \cdot \vec{S}_{i+1}\n
&& -h\sum_{i} S^z_i + \sum_i V(\delta_{i})
\label{eq:ham}
\ea
Here $\delta_i$ denotes the interatomic distance between site $i$ and $i+1$, $h$ is external magnetic field and
$\vec{S}_{i}$ are spin $1/2$ operators.

The dependence of the spin-phonon coupling on the interatomic distance $\delta_i$
has been expanded up to second order with coefficient
$A_2$. A Zeeman term is included to take into account magnetic field effects.

The phonon potential energy in (\ref{eq:ham}) is given by
\be V(\delta_{i}) = \omega_0 \left( \frac{1}{2} \delta_{i}^2 + \alpha_3
\delta_{i}^3 + \alpha_4 \delta_{i}^4 \right) \ee
where $\alpha_{3}$ and $\alpha_{4}$ take into account the anharmonicity of the
interatomic potential energy.

Generally, the properties due to the anharmonic oscillations arise both from the addition of
quartic terms in the potential energy and next-to leading terms in the
spin-phonon coupling. In this letter we focus on the contribution of
the anharmonicity in the spin-phonon coupling measured by $A_2$ ignoring the contribution
of higher-order terms in the potential energy. We show below that it is the term quadratic
in the lattice deformations in the interaction Hamiltonian that changes drastically
the physics of the magnetoelastic $XXZ$ chain. We expect that higher order terms
in the potential energy (cubic and quartic) are inessential.


\section{Bosonization description}

Following the usual procedure in the low energy limit, we bosonize
the spin degrees of freedom at fixed magnetization $M$ and the interaction term becomes
\cite{us}
\be
H_{sp-ph}=\int dx \left( A_1\,\delta_M(x)+A_2\, \delta_M(x)^2 \right)
\rho(x)
\label{intbos}
\ee
where we have introduced the subscript $M$ in $\delta_M(x)$ to stress its
dependence on the magnetization. Here $\rho(x)$ is the continuum expression of the energy density
\be
\rho(x) =\alpha  \partial_x \phi
 + \beta  \cos(2k_F x+\sqrt{2\pi}\phi)  + \cdots
\label{robos}
\ee
where $k_F = \frac{\pi}{2}(1-M)$, $\alpha$ and $\beta$ are
constants and the ellipses indicate higher harmonics
\cite{Haldane80}.

The main contribution in the low energy limit comes from the
constructive interference between the modulation term
$A_1\delta_M(x)+A_2 \delta_M(x)^2$ and the most relevant part of
$\rho(x)$, {\it i.e.} $\cos(2k_F x+\sqrt{2\pi}\phi)$.  This operator has conformal dimension that depends on the Tomonaga-Luttinger parameter $K(M,\Delta)/2$ where $\Delta$ measures the z-axis anisotropy in the $XXZ$ model. Here we emphasize the dependence on the magnetization $M$ and the anisotropy $\Delta$.

Let us propose a periodic pattern of deformations $\delta_{M}(x)$
with period $L_p \ $ ,  {\it i.e.} satisfying $\delta_{M}(x+L_p)=\delta_{M}(x)$
(the lattice spacing $a$ is set to $1$ in what follows, so that $L_p$ is an integer).
The most general Ansatz for the modulation
term is given by
\ba
\delta_{M}(x)=
\sum_{n=1}^{N_w} \delta_{n}(M)\cos\left( n\frac{2\,\pi\,x}{L_p}+\theta_n(M)\right)
\label{delta(M)}
\ea
where  $\delta_{n}(M)$ are the amplitudes and $\theta_n(M)$  the phases of the different terms in the expansion.
The upper sum index $N_w$ equals $L_p/2$ if $L_p$ is even and $(L_p-1)/2$ if it is odd. (In what follows the dependence of $\delta_{n}(M)$ and $\theta_n(M)$ on $M$ is suppressed to ease the notation, i.e. $\delta_{n}(M)\to \delta_{n}$ and $\theta_n(M)\to\theta_n$).

From Eqs.\ (\ref{robos}) and (\ref{delta(M)}), we see that the product between the
two terms is commensurate whenever the following relation is satisfied

\be
k_F \propto \frac{2\,\pi}{L_p} ,
\label{6}
\ee
which implies that the wavelengths of the modulations that could pin
the relevant cosine term are related to the magnetization as

\be
L_p=\frac{4\,m}{1-M} ,
\label{7}
\ee
where $M \neq 1$ and $m$ an arbitrary integer, the smallest possible that makes $L_p$ an integer.

%
%

The Ansatz in Eq.\ (\ref{delta(M)}) is verified a posteriori from the DMRG analysis, where it is seen that the
modulation amplitudes $\delta_n$ and the phases $\theta_n$ depend strongly on the value of
the magnetization $M$, some of them being zero in certain cases.

Using this form for $\delta_M(x)$, the interaction term (\ref{intbos}) takes the form
\be
H_{sp-ph}=\sum_{p=0}^{2(N_w+1)} \,\lambda_p \,\int dx \,\cos(p\,k_F x+\sqrt{2\pi}\,\phi+\Gamma_p)
\label{H:sp-ph}
\ee
where $\Gamma_p$ is a function of the phases ${\theta_n}$ in the expansion (\ref{delta(M)}), and $\lambda_p$ is a function
of ${\delta_n}$,  ${\theta_n}$ and the coupling constants $A_1$ and $A_2$.

This form of the interaction allows us to conclude that the spin
Peierls effect takes place in the usual manner (see \cite{us} and references
therein), since we have both the always commensurate term ($p=0$ in the above
equation)  $\cos(\sqrt{2\pi}\,\phi)$ and the $4k_F$ term, that provide together
a dimerization of the lattice and a plateau at $M=0$ in the magnetization curve.

For finite magnetization, using Eqs.\ (\ref{6})-(\ref{H:sp-ph}) and using the commensurability
condition that arises from (\ref{H:sp-ph}), $p\,k_F/2\pi \in Z$, one obtains the following
condition for the frequencies in (\ref{delta(M)}) to pin a relevant perturbation

\ba
(z\pm 2)\,(1-M)&=&4\times\text{integer}
\label{necessary-condition}
\ea
where $z$ is an integer that runs through all the frequencies that
appear in the lattice deformation Eq.\ (\ref{delta(M)}) and its square, i.e. $z=0,...,2\,N_w$. In Table \ref{table-frecuencies}
we show some examples that we analyze in what follows using DMRG.
\begin{table}
\begin{center}

\begin{tabular}{|c|c|c|c|}
\hline
&$M=1/5$ &$M=1/3$ &$M=1/2$\\
\hline
$L_p$& $5$ &$6$&$8$\\
\hline
$N_w$& $2$ &$3$&$4$\\
\hline
$z\pm 2$&$5,10,...$&$6,12,...$&$8,16,...$ \\
\hline
$z$&$3$&$4$&$6$ \\
\hline
possible &                &$2k_F$&\\
         &$k_F$ and $2k_F$&or&$2k_F$ and $4k_F$\\
frequencies&&$k_F$ and $3k_F$&\\
\hline
\end{tabular}
\end{center}
\caption{Possible frequencies for the lattice deformations for
magnetizations $M=1/5$, $1/3$ and $1/2$, obtained from the Eq.\ (\ref{necessary-condition}).}
\label{table-frecuencies}
\end{table}

One should stress that in the present case the situation is rather different
than in previous studies of spin systems in a magnetic field, such
as in the case of spin ladders, magnetoelastic zig-zag chains, etc., since now
the perturbing operator that would be responsible for the plateau
is relevant, independently of the values of the microscopic parameters. This
may seem to imply that condition (\ref{necessary-condition}) is also sufficient, but
bosonization alone does not provide the actual values of the amplitudes
of the different Fourier components of the deformation we proposed
and it remains to be checked that they are indeed non vanishing.
In order to answer this question we need to use DMRG as we describe below.

From the above analysis, we predict that the magnetization curve may present
new features related to the
frequencies which appear in the Fourier decomposition of the elastic deformation (\ref{delta(M)})
for some given values of $M$, such as $M=1/5$, $M=1/3$, $M=1/2$, etc.
Since these frequencies pin the very relevant term
$\cos(\sqrt{2\pi}\,\phi)$, plateaux at these values of $M$ are expected to show up even for a
small anharmonicity $A_2$. In such cases, the plateaux widths $Gap(M,\Delta)$ should scale as

\be
Gap(M,\Delta) \propto \lambda^{1/(2-d(M,\Delta))}
\label{scaling}
\ee
where $\lambda$ is the coupling constant associated to the relevant cosine term in
$H_{sp-ph}$ and $d(M,\Delta)$ is the scaling dimension which
can be computed from the Bethe Ansatz solution \cite{CHP}. The coupling $\lambda$ is a function of the
anharmonic amplitude $A_2$ and its functional dependence though not predicted by bosonization,
can be computed numerically as we show below. From now on we will concentrate in the isotropic case $\Delta=1$.

\section{DMRG analysis}

This is the general setting obtained from bosonization which provides the qualitative picture
expected when anharmonic effects play a role. To have a complete and more quantitative picture
we study the system using extensive DMRG computations.
More specifically, we compute the ground state energy $E(S^{z}_{total},h=0)$ of Eq.~(\ref{eq:ham}) in the complete set
of $S^{z}_{total}$ subspaces using periodic boundary conditions, and keeping just 300 states it was
shown to be enough to assure the accuracy of the calculation. As usual, adding the Zeeman term, we solve
the equation $E(S^{z}_{total},h) = E(S^{z}_{total}+1,h)$ to obtain the normalized magnetization $M=2S_z/N$
where the plateaux are showing up.
This procedure allows us to compute the actual width of the plateaux and their scaling behavior, the
deformation patterns and fractional excitations for the different plateaux.

\begin{figure}[t]
 \begin{center}
 \includegraphics[height=6.1cm]{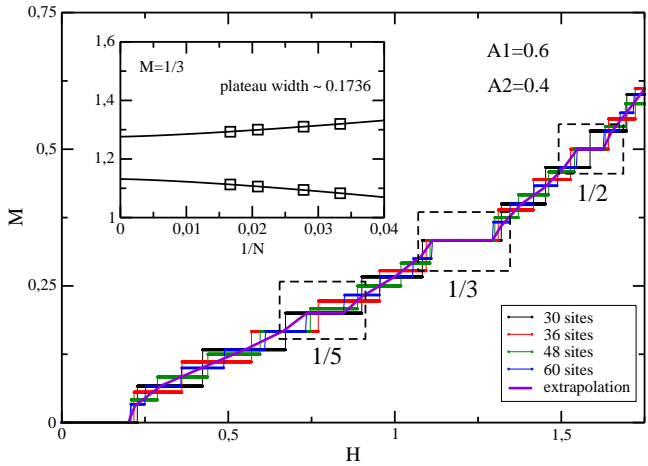}
 \end{center}
\caption{M vs. h for $A_1=0.6, A_2=0.4$ and different system sizes.
($N = 30, 36, 48$ and $60$). The bold purple line corresponds to the extrapolation
to the thermodynamic limit. The plateaux at $M=0$ and $1/3$ are clearly observed,
while for $M= 1/5$ and $1/2$, it is hard to conclude
if they survive in the thermodynamic limit. Note that for $N=30$, $M=1/5$, $1/2$ are not commensurate.
The inset shows the finite size scaling of the width of the  plateau at $M=1/3$. Its finite size scaling is
expected to follow $width(N) = width(\infty) + A\, N^{-B}$.}
\label{fig:f1}
\end{figure}

Let us analyze in detail the situation at $M=1/3$, where we expect to have a plateau.
In this case $k_F=\pi/3$ and our Ansatz for the modulation (\ref{delta(M)}) takes the form
\begin{eqnarray}
\nonumber \delta_{1/3}(x)&=&\dk\, \cos(k_F\,x+\theta_{1})+\dkk\, \cos(2k_F\,x+\theta_{2})\\
&+&\dkkk\, \cos(3k_F\,x+\theta_{3})
\label{deltaM1on3}
\end{eqnarray}
which leads to the perturbation Hamiltonian

\begin{eqnarray}
H_{sp-ph} \approx \lambda_{1/3}\ \cos\left(\sqrt{2\pi}\,\phi+\Gamma_{1/3} \right) + \cdots
\label{Hint-1}
\end{eqnarray}
where $\lambda_{1/3}$ and $\Gamma_{1/3}$ depend on $\lambda_{0}$ and $\lambda_{6}$ which are the
only two commensurate terms in Eq.~(\ref{H:sp-ph}) at magnetization $M=1/3$ (see Fig.~\ref{fig:f1}).
The dots in the equation above indicate less relevant terms,
which can be safely discarded in the presence of the more relevant term
$\propto \cos\left(\sqrt{2\pi}\,\phi \right)$.

The couplings appearing in (\ref{Hint-1}) have a lengthy expression in terms of the strengths of the spin-phonon couplings
$A_1$ and $A_2$ but also on the $\dnk$'s and on the relative phases $\theta_{n}$'s, whose values
cannot be extracted from the bosonization analysis alone. To further proceed we now resort to the
numerical analysis of the system using DMRG on large systems which allows us to estimate all these
parameters in a self-consistent way.

The lattice deformations can be calculated in a self consistent way. Minimizing the ground state energy
and imposing the following constraint
\ba
\sum_{j}\delta_{j}=0
\ea
we obtain
%
\ba
\nonumber
\delta_{i}=\frac{JA_{1}
\left[
\left( \frac{\sum_{k}\langle \vec{S}_{k} \cdot \vec{S}_{k+1}\rangle\;
(\omega_{0}+2JA_{2}\langle \vec{S}_{k} \cdot \vec{S}_{k+1}\rangle)^{-1}}{\sum_{k}
(\omega_{0}+2JA_{2}\langle \vec{S}_{k}
 \cdot \vec{S}_{k+1}\rangle)^{-1}} \right)
-\langle \vec{S}_{i} \cdot \vec{S}_{i+1}\rangle
\right]
}
{(\omega_{0}+ 2JA_{2}\langle \vec{S}_{i} \cdot \vec{S}_{i+1}\rangle)}\\\label{eq:selfconsistent}
\ea
%
We start from  an arbitrary
chosen initial set of deformations $\{\delta^{(0)}\}$ to be varied and determined self consistently. For a given
set $\{\delta^{(N)}\}$  we determine
the corresponding ground state and then we compute a new set $\{\delta^{(N+1)}\}$ using
 (\ref{eq:selfconsistent}) which we use again in the Hamiltonian. Iterating this procedure,
 we finally obtain a fixed point configuration of the deformations
 $\delta_{i}^{(N+1)}(\{\delta^{(N)}\})=\delta^{(N)}_{i}$.

\begin{figure}[htb]
 \begin{center}
\includegraphics[height=6.7cm]{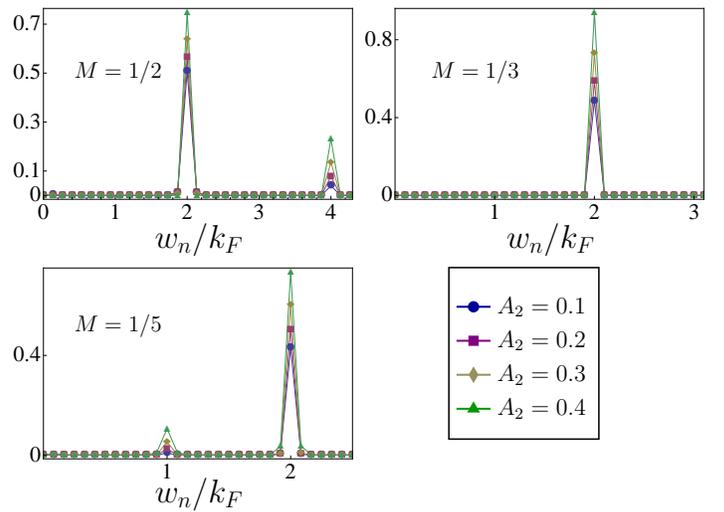}\end{center}
\caption{Amplitudes $\delta_n(M)$ (see Eq.\ (\ref{delta(M)}))) as a function of the frequency $w_n=2\pi\,n/L_p$
in units of $k_F$ (with $A_1=0.6$) for $M= 1/2, 1/3$ and $1/5$. The peaks indicate which frequencies
contribute to the deformation pattern. The $2\,k_F$ peak is always bigger because to linear order
in $\delta_M(x)$ it contributes to the energy for all magnetizations.}
\label{fig:f2}
\end{figure}

From the DMRG data, we observe that for $M=1/3$  only the $2k_F$ mode contributes to
the lattice deformations (see Fig.~\ref{fig:f2}), so that we can safely set $\dk=\dkkk=0$. As for
the phase $\theta_2$, it is negligible within the numerical precision so we set it to zero in
what follows. With this input from DMRG, we get the following expressions for the bosonization parameters,
{\it i.e.} for the amplitude $\lambda_{1/3}$ and phase $\Gamma_{1/3}$ in Eq.\ (\ref{Hint-1}),
%
%
\begin{eqnarray}
\lambda_{1/3}&\propto&\left(\frac{A_1\,\dkk}{\sqrt{8}}+\frac{A_2\,\dkk^2}{\sqrt{32}}\right)\nn\\
\Gamma_{1/3}&=&-\pi/3
\label{eq:lambdaytheta}
\end{eqnarray}

Here a word is in order: To analyze the scaling of the gap we need to identify the effective coupling constant
associated to the perturbing operator responsible for the opening of the gap. Since the term
proportional to $A_1$ is present for all magnetizations, it does not play a role in the gap
opening and we can then identify the coupling constant governing the scaling of the gap
in (\ref{scaling}) as $\lambda \propto A_2\,\dkk^2$.

On the other hand, we can extract the deformation amplitude as a function of $A_2$ from
the numerical data, which after a finite size scaling analysis and a square fit leads to $\dkk=a+b\,A_2+c\,A_2^2$
with $a=0.110$, $b=0.098$ and $c=0.551$. Now that we have the dependence
of the effective coupling $\lambda_{1/3}$ on the anharmonicity $A_2$
we can analyze the scaling of the spin gap (the width of the plateau), which should scale as
in (\ref{scaling}).

\begin{figure}[htb]
\centering
\includegraphics[width=0.47\textwidth]{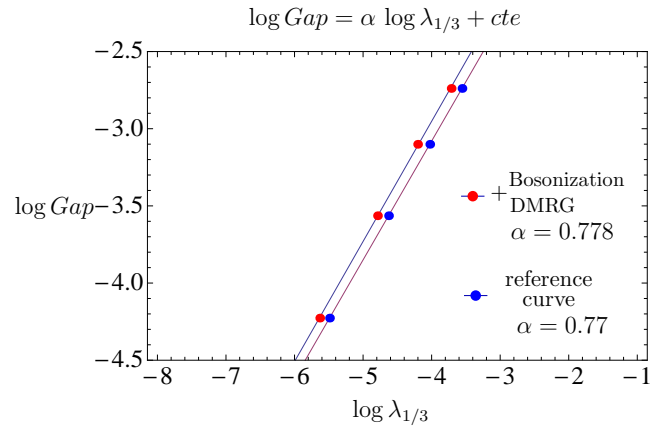}
\caption{Logarithmic plot of the $Gap(\lambda)$: The blue dots correspond to the reference
curve $Gap(\lambda)=\lambda^{0.77}$ with the gap
obtained from DMRG, while the red dots correspond to the gap obtained from
DMRG vs. the values of $\lambda$ extracted from bosonization. The value $0.77$ is
obtained from the Bethe Ansatz solution}
\label{fig:f3}
\end{figure}

In order to compare both approaches, we need to use the relation (\ref{eq:lambdaytheta}) between $\lambda_{1/3}$ and $A_2$, together with the values of $\dkk$ obtained form DMRG. Following this
approach, in  Fig.~\ref{fig:f3} we show a logarithmic plot of the gap vs $\lambda_{1/3}$ using the values of $\lambda_{1/3}$ obtained by bosonization and those of the gap by DMRG (red points). We show a linear fit to obtain the exponent in eq.\ (\ref{scaling}) and compare with a reference line (blue points) to show the agreement of both approaches.
%

\section{Conclusions}

In conclusion, we have described a new mechanism leading to the formation of rational magnetization plateau phases, which is mainly due to the anharmonic spin-phonon coupling. We have shown that its role is to pin magneto-elastic deformations which are not present in the harmonic case. By means of bosonization we have shown that the inclusion of the anharmonic spin-phonon coupling gives as a contribution a relevant operator that is responsible for the plateau in the magnetization curve for certain commensurate values of the magnetization $M$. We have performed extensive DMRG computations to complement the analytical computations, since the bosonization approach alone does not provide the actual values of the amplitudes of the different Fourier components of the lattice deformations.  In particular, we have analyzed in detail the situation at $M=1/3$ where we have computed the plateau width as a function of the anharmonic coupling, to extract the  scaling dimension of the relevant operator that opens the gap. Finally, we have seen that the exponent obtained from the DMRG computations and the one obtained from the Bethe Ansatz through bosonization, are in excellent agreement, providing further support to our results.

\noindent Acknowledgements: This work was partially supported by the ESF through INSTANS, PICT
ANPCYT Grant No 20350, and PIP CONICET Grant No 1691. C.J.G acknowledges support from PICT
ANPCYT Grant No 1647 and No 1776, and PIP CONICET Grant No 0392.

\end{document}